\RequirePackage{shorthand}
\documentclass{moriond}

\def\be{\begin{equation}}
\def\ee{\end{equation}}
\def\bea{\begin{eqnarray}}
\def\eea{\end{eqnarray}}

\newcommand{\dmu}{\partial_\mu}

\newcommand{\hmn}{\widehat{F}_{\mu\nu}}
\newcommand{\hmnup}{\widehat{F}^{\mu\nu}}

\newcommand{\half}{\frac{1}{2}}

\newcommand{\Photo}{}

\begin{document}
\vspace*{4cm}
\title{THE 750 GEV EXCESS FROM PHOTON-PHOTON AND QUARK-QUARK PROCESSES}

\author{TANUMOY MANDAL, ULF DANIELSSON, RIKARD ENBERG, GUNNAR INGELMAN}

\address{Department of Physics and Astronomy, Uppsala University, Box 516, SE-751 20 Uppsala, Sweden}

\maketitle\abstracts{
The observed excess in the diphoton mass spectrum around 750 GeV at the 13 TeV LHC possibly indicates the presence of a photonphilic resonance.   
We show that the excess can be explained by a scalar of the type involved in Bekenstein's framework for varying electromagnetic coupling theories. The scalar, in our
model, couples dominantly to photons and is mainly produced by the quark-quark fusion at the LHC.  
In addition, it can also be produced in photon-photon fusion. Our model has only
two free parameters, the mass of the scalar and the scale of the new physics, which are fixed by the LHC excess to 750 GeV and 1.5 - 2 TeV, respectively. The scalar has a large three-body decay to a fermion pair and a photon, which provides an interesting search channel with a dilepton-photon resonance around 750 GeV.}

\section{Introduction}

Recently both the ATLAS and the CMS collaborations have reported excesses in
the diphoton invariant mass distribution around 750 GeV~\cite{ATLAS:2015dxe,CMS:2015dxe} based on the 13 TeV data from run-II. The excess is only observed in the
diphoton channel with possibly no other hard activities in association with the
resonance. There is also a tension
between the excess observed in run-II data and non-observation of any such excess
in run-I data. This tension depends on how the resonance is produced. For instance, if the resonance is dominantly produced from the photon-fusion, the tension is more compared to the gluon-fusion production. But due to large uncertainties involved in the photon-flux
from proton, run-I data might be compatible to the run-II data.  
We show that the excess can be explained by a scalar of the type involved in Bekenstein's framework for varying electromagnetic coupling theories simultaneously 
satisfying all the relevant run-I data. Although the excess is not yet statistically significant enough to draw any definite conclusion, but are nevertheless very interesting to understand from a theoretical perspective. In this talk, we propose to explain the observed excess by a massive scalar associated with a model for a space-time varying electromagnetic coupling constant, $e$ and the talk is mainly based on our paper in Ref.~\cite{Danielsson:2016nyy} on this subject.
The original concept of space-time varying fine-structure constant was constructed by Jacob Bekenstein~\cite{Bekenstein:1982eu} in context of cosmology. 

\section{The Model}

In the Bekenstein model~\cite{Bekenstein:1982eu}, it is assumed that $e$ is not really a constant and varies with space-time. The variation is given by $e = e_0 \epsilon(x)$, where $e_0$ is the constant EM coupling and $\epsilon(x)$ is a scalar field whose kinetic energy is given by
$\Lambda^2(\partial_\mu\epsilon)^2/(2\epsilon^2)$
where $\Lambda$ is an energy scale. The field strength tensor is given by
$F_{\mu\nu} = \left[ \partial_\mu (\epsilon A_\nu) - \partial_\nu (\epsilon A_\mu) \right]/\ep$.
We define 
$\hmn=\partial_{\mu} A_\nu - \partial_\nu A_\mu$, and introduce a scalar field $\varphi$ such that $\epsilon = e^\varphi$. Assuming small field limit, we write  $\epsilon \simeq 1+\varphi$, and keep only terms linear in $\varphi$. Finally, we define a new field $\phi=\varphi \Lambda$, so that all fields have their usual mass dimensions. In this way, we find a Lagrangian for standard electromagnetism plus the terms associated with the scalar field as
\begin{equation}
 {\cal L} \supset \half(\dmu\phi)^2
-\frac{1}{2}M_\phi^2\phi^2  + \frac{1}{2\Lambda}\phi\hmn\hmnup .
\label{eq:scalarlagr}
\end{equation}

\section{Production and decays at the LHC}

We study the LHC phenomenology of our model and derive limits on the
scale $\Lambda$ from the relevant 8 TeV data and also find out the favored values of $\Lambda$
required to explain the 750 GeV diphoton excess.

\begin{figure}[!ht]
\centering
\includegraphics[scale=0.5]{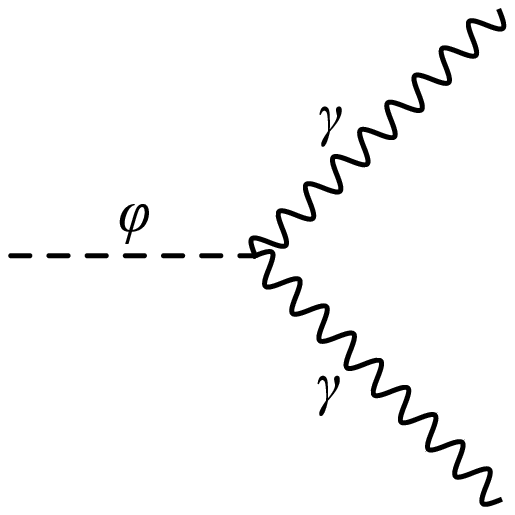}~~~~~
\includegraphics[scale=0.5]{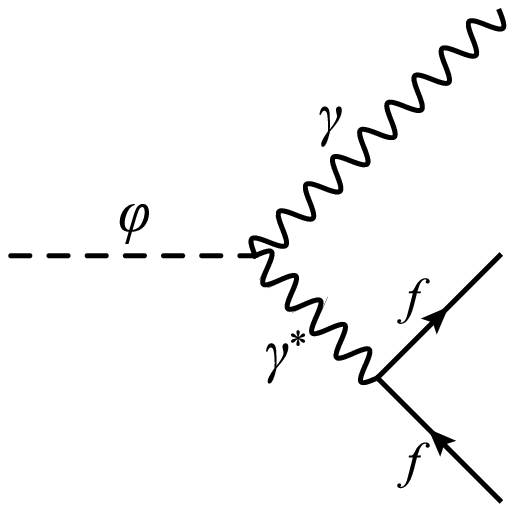}~~~~~
\includegraphics[scale=0.5]{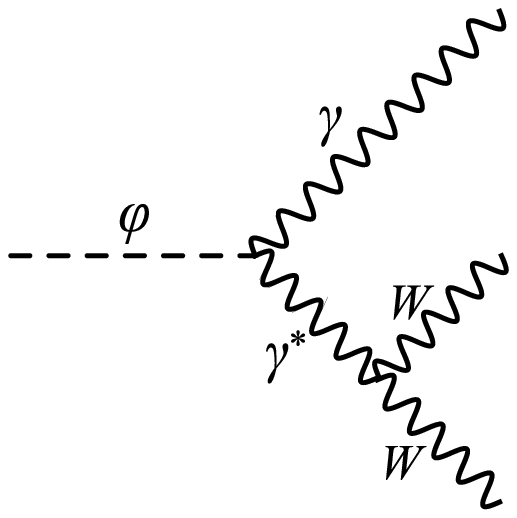}
\caption{Sample Feynman diagrams of the two and three body decay modes of $\phi$.}
\label{fig:FDhpDecay}
\end{figure}

The only possible
two-body decay of $\phi$ is the diphoton mode, which is a tree level decay -- not a loop-induced
decay through a charged particle. In Fig.~\ref{fig:FDhpDecay},
we show the Feynman diagrams of all possible two and three body decay modes of $\phi$. 
In Table~\ref{tab:PWBR}, we show the partial widths and branching
ratios (BR) of $\phi$ into its two, three and four body decay modes for $\Lambda=1$ TeV, calculated using 
{\sc MadGraph5}.
It is important to note that the branching ratios of $\phi$ are independent of $\Lambda$.
From Table~\ref{tab:PWBR}, we can see that $\phi\to\gamma\gamma$ is the dominant decay mode and this mode has a branching ratio
of about 70\%. Due to the large branching in the diphoton mode, $\phi$ might be a good candidate to explain the
recent 750 GeV diphoton excess at the LHC. Moreover, the nonobservation of any excess expected in other channels from a SM-like
scalar can also be explained as other decays of $\phi$ are not SM-like. 
\begin{table}[!ht]
\centering
\begin{tabular}{|c|c|c|c|c|c|c|}
\hline
Decay Mode & $\phi\to \gamma\gamma$ & $\phi\to \gamma ff(jj)$ & $\phi\to \gamma\gamma ff$ & $\phi\to \gamma WW$ & $\phi\to \gamma\gamma WW$ & Total \\
\hline
Width (GeV) & 8.393 & 2.672 (1.505) & 0.610 & 0.447 & 0.022 & 12.14 \\
\hline
BR (\%) & 69.12 & 22.00 (12.39) & 5.02 & 3.68 & 0.18 & - \\
\hline
\end{tabular}
\caption{The partial widths and branching ratios of $\phi$ for $M_{\phi}=750$ GeV. The widths are proportional to $\Lambda^{-2}$ and are here given for $\Lambda=1$ TeV, whereas the BRs are independent
of $\Lambda$. Here, $f$ includes all SM charged fermions and $j$
denotes jets of ``light" quarks, including $b$.}
\label{tab:PWBR}
\end{table}

\begin{figure}[!ht]
\centering
\includegraphics[height=3cm,width=3cm]{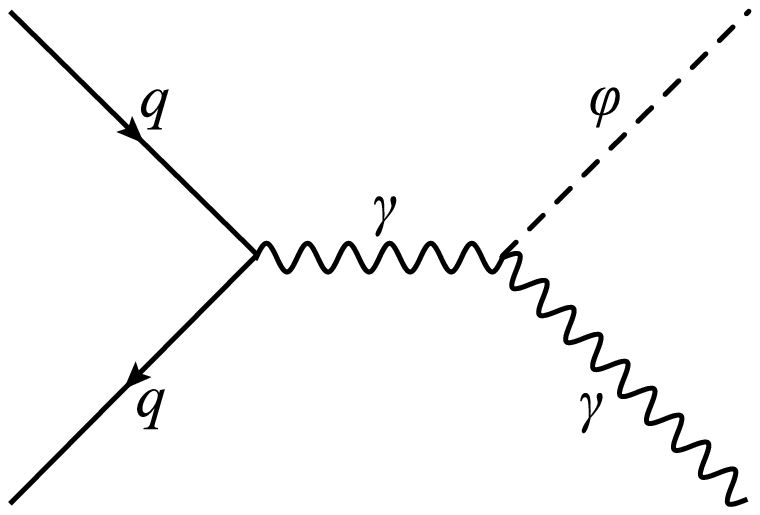}~~~~~
\includegraphics[height=3cm,width=3cm]{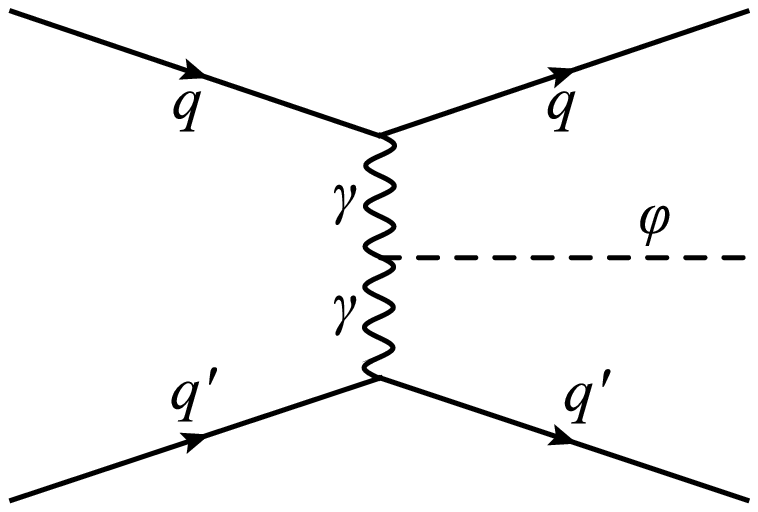}
\caption{Sample Feynman diagrams of the production of $\phi$ at the LHC.}
\label{fig:FDpp2hpx}
\end{figure}
\begin{table}[!ht]
\centering
\begin{tabular}{|c|c|c|c|c|c|c|c|c|c|}
\hline
Production mode & $\phi\gamma$ & $\phi\gamma j$ & $\phi\gamma jj$ & $\phi jj$ & $\gamma\gamma\to\phi$ & $\phi\ell\ell$ & $\phi\gamma W$ & $\phi\gamma WW$ & $\phi\gamma\gamma$ \\
\hline
CS@8TeV (fb)  & 2.339 & 1.161 & 0.513 & 4.622 & 30.84 & 0.017 & 0.064 & 0.176 & 0.026 \\
\hline
CS@13TeV (fb) & 10.63 & 6.486 & 3.439 & 15.15 & 77.59 & 0.081 & 0.577 & 4.847 & 0.140 \\
\hline
\end{tabular}
\caption{Partonic cross sections of various production channels of $\phi$ for $\Lambda=1$ TeV computed at renormalization ($\mu_R$)
and factorization ($\mu_F$) scales $\mu_R=\mu_F=M_{\phi}=750$ GeV for LHC at 8 and 13 TeV. These cross
sections are computed using CTEQ6L1 PDFs by applying some basic generation level cuts.
Here, $j$ denotes light jets including $b$-jet and $\ell$ includes $e^{\pm}$ and $\mu^{\pm}$. In the $\gamma\gamma\to\phi$ mode, the
two initial photons come from proton.}
\label{tab:CS}
\end{table}
In Fig.~\ref{fig:FDpp2hpx}, we show a few sample Feynman diagrams of the two main production
channels of $\phi$ at the LHC.
Unlike the $gg$ initiated SM-like Higgs boson production, these channels are
induced by a $qq$ initial state. In Table~\ref{tab:CS}, we present the partonic cross sections of
various production modes of $\phi$ for the 8 and 13 TeV LHC for $\Lambda=1$ TeV.

The experimental searches for a high mass diphoton resonance, where the signal is an $s$-channel
spin-0 or spin-2 resonance decaying to diphoton generally demand exactly two selected photons with no selected jet, whereas for inclusive diphoton resonance searches, one keeps events with at least two selected photons
and any number of selected jets. The two important production channels of $\phi$ {\it viz.}
inclusive $pp\to\phi\gamma\to 3\gamma$ and $pp\to\phi jj\to\gamma\gamma jj$ contribute most to the ATLAS~\cite{ATLAS:2015dxe} and CMS~\cite{CMS:2015dxe} analyses.
Now, we want to investigate how selection cut efficiencies depend on the different selection 
criteria imposed on the number of photons and jets.
In Table~\ref{tab:CEhpa}, we show cut efficiencies for different selection
criteria on the number of photons and jets for the inclusive $pp\to\phi\gamma\to 3\gamma$ and $pp\to\phi jj\to\gamma\gamma jj$ channels at the 13 TeV LHC by
roughly employing the selection cuts used by ATLAS~\cite{ATLAS:2015dxe} and CMS~\cite{CMS:2015dxe}
for their 13 TeV analyses. 

\begin{table}[!ht]
\centering
\begin{tabular}{|c|c|c|c|c|c|c|c|}
\hline
Category & $2\gamma+0j$ & $\geq 2\gamma+0j$ & $2\gamma+\geq 0j$ & $2\gamma+1j$ & $2\gamma+2j$ & $3\gamma+\geq 0j$ & $\geq 2\gamma+\geq 0j$\\
\hline
\hline
ATLAS ($\phi\gamma$) & 0.008 & 0.278 & 0.142 & 0.057 & 0.042 & 0.641 & 0.788 \\
\hline
ATLAS ($\phi jj$) & 0.0009 & 0.001 & 0.546 & 0.030 & 0.219 & 0.010 & 0.556 \\
\hline
\hline
CMS ($\phi\gamma$) & 0.036 & 0.323 & 0.247 & 0.086 & 0.063 & 0.704 & 0.957 \\
\hline
CMS ($\phi jj$) & 0.0009 & 0.001 & 0.750 & 0.035 & 0.291 & 0.013 & 0.763 \\
\hline
\end{tabular}
\caption{Cut efficiencies for different selection criteria on the number of selected photons and jets for the 13 TeV ATLAS and CMS 
diphoton resonance searches. Here, ``$\phi\gamma$'' and ``$\phi jj$'' mean inclusive (up to 2-jets) $pp\to \phi\gamma\to 3\gamma$ and $pp\to \phi jj\to \gamma\gamma jj$ processes respectively.}
\label{tab:CEhpa}
\end{table}

In order to derive a limit on $\Lm$ by recasting the $\sg\times{\rm BR}$ upper limit from an experiment, we need to properly take care
of the cut efficiencies. This can be done by using the following relation:
\be
\label{eq:Ns}
\mc{N}_s =\sg_s\times \ep_s\times \mc{L}=\lt(\sum_i \sg_i\times\ep_i\rt)\times\mc{L}\ ,
\ee
where $\mc{N}_s$ is the number of events for the signal considered and $\sg_s$ is the
corresponding signal cross section for luminosity $\mc{L}$. The corresponding signal
cut efficiency is denoted by $\ep_s$. When different types of signal topology and/or
final state contribute to any experimental observable, $\mc{N}_s$ can be expressed by the sum
$\lt(\sum_i \sg_i\times\ep_i\rt)\times\mc{L}$. Here, $i$ runs over all contributing
signal processes to any observable. In our case, the following two processes contribute the most
to the $s$-channel diphoton resonance searches at the LHC:
\be
{\rm Process~I}~(p_1):~pp\to \phi\gm\to 3\gm+{\rm jets};~~{\rm Process~II}~(p_2):~pp\to \phi jj\to \gm\gm jj .
\ee
Therefore, in our case $\mc{N}_s=\Lm^{-2}\times(\sg_{p_1}\times\ep_{p_1}+\sg_{p_2}\times\ep_{p_2})_{\Lm=1~{\rm TeV}}\times\mc{L}$.
Here, $\sg_{p_i}$ is the signal cross section for the
$i$-th process and the corresponding signal efficiency is $\ep_{p_i}$. 

\begin{table}[t]
\centering
\begin{tabular}{|c|c|c|c|c|c|c|c|c|c|c|}
\hline
Experiment & $\sg_s$ & $\ep_s$ & $\sg_{p_1}$ & $\ep_{p_1}$ & $\ep_{p_1}$ & $\sg_{p_2}$ & $\ep_{p_2}$ &
 $\ep_{p_2}$ & $\Lm$ (C1) & $\Lm$ (C2) \\
& (fb) & & (fb) & (C1) & (C2) & (fb) & (C1) & (C2) & (TeV) & (TeV) \\
\hline
ATLAS@13TeV & 10.5 & 0.4 & 9.9 & 0.14 & 0.79 & 9.7 & 0.55 & 0.56 & 1.3 & 1.8 \\
\hline
CMS@13TeV & 17.0 & 0.3 & 9.9 & 0.25 & 0.96 & 9.7 & 0.75 & 0.76 & 1.4 & 1.8 \\
\hline
\end{tabular}
\caption{The observed upper limit on cross sections, $\sg_s$ and corresponding efficiencies, $\ep_s$ for mass around 750 GeV. In the last two columns we show the derived value of $\Lm$ for $M_{\phi}=750$ GeV for the selection categories C1: $2\gm+\geq 0j$ and C2: $\geq 2\gm+\geq 0j$.}
\label{tab:ltF}
\end{table}

The LHC Run-I data of ATLAS and CMS set an upper limit of the cross section in the range 1-2 fb. Using this we can extract a lower limit of $\Lm$ for our model. For $2\gm + 0j$ category, we obtain a lower limit of $\Lm$ in the range 0.2-0.6 TeV. Choosing instead $2\gm + \geq 0j$ would, however, give a lower limit $\Lm$ in the range 1.2-1.8 TeV.
From the 13 TeV data, where a more pronounced hint for an excess can give more precise information regarding our model. We obtain the essential results shown in Table~\ref{tab:ltF}, resulting in 
values of $\Lm$ that can
explain the 750 GeV diphoton excesses observed by both ATLAS~\cite{ATLAS:2015dxe}
and CMS~\cite{CMS:2015dxe}. As we mentioned earlier, the extraction
of $\Lm$ depends on what selection category is used. For the
category of $2\gm + 0j$ (C1) selection for ATLAS analysis, we get $\Lm\approx 1.3$ TeV which can explain the excess. On the other hand we get slightly bigger $\Lm\approx 1.8$ TeV for the selection category
$\geq 2\gm + \geq 0j$ (C2). The corresponding CMS values on $\Lm$ are 1.4 TeV and 1.8 TeV
for C1 and C2 respectively.

\section{Conclusions and Outlook}

We have proposed that the diphoton excess is due to a 750 GeV scalar associated with variations of the fine-structure constant. Our model has only two new parameters, the mass of the scalar $M_\phi$ and the energy scale $\Lambda$. Both are fixed by the LHC excess, with $M_{\phi}\sim 750$~GeV and $\Lambda\sim 1.5-2$~TeV.

Our proposal predicts that
the scalar dominantly decays to a photon pair, but has an appreciable branching ratio into a pair of fermions plus a photon, with BR($\phi\to\gamma q\bar q)\sim 13\%$ for quarks and BR($\phi\to\gamma\ell^+\ell^-)\sim 10\%$ for leptons. This gives the main additional prediction of our model: events with a lepton-lepton-photon resonance at 750 GeV.
The scalar resonance is dominantly produced together with an additional real or virtual photon, which, if virtual, gives rise to a pair of jets or leptons. This gives another prediction: the existence of an additional photon and/or jets in the events, which are not part of the resonance. These predictions should be looked for in future LHC analyses.

The scalar can be produced through photon-photon fusion where the initial photons are 
coming from proton and this production can be very large due to the infra-red
enhancement in the collinear limit. Due to the limited knowledge of proton form factors,
the photon fusion contribution inherits large uncertainties~\cite{Harland-Lang:2016qjy}. Therefore, it is very important to understand the photon PDF in proton for robust predictions. Currently, we
are working on this direction to understand various issues in the photon-flux to improve
our model predictions.

\section*{Acknowledgments}

TM would like to thank the organizers of the Moriond QCD 2016 for giving the opportunity to
present this result.
This work is supported by the Swedish Research Council under contracts 621-2011-5107 and
2015-04814. T.M. is supported by the Carl Trygger Foundation under contract CTS-14:206.


\section*{References}

\begin{thebibliography}{99}

\bibitem{ATLAS:2015dxe} 
  The ATLAS collaboration,
  ATLAS-CONF-2015-081.
  
\bibitem{CMS:2015dxe} 
  CMS Collaboration [CMS Collaboration],
  collisions at 13TeV,''
  CMS-PAS-EXO-15-004.
  
\bibitem{Danielsson:2016nyy} 
  U.~Danielsson, R.~Enberg, G.~Ingelman and T.~Mandal,
  arXiv:1601.00624 [hep-ph].
  
\bibitem{Bekenstein:1982eu} 
  J.~D.~Bekenstein,
  Phys.\ Rev.\ D {\bf 25}, 1527 (1982).
  doi:10.1103/PhysRevD.25.1527
  
\bibitem{Harland-Lang:2016qjy} 
  L.~A.~Harland-Lang, V.~A.~Khoze and M.~G.~Ryskin,
  JHEP {\bf 1603}, 182 (2016)
  doi:10.1007/JHEP03(2016)182
  [arXiv:1601.07187 [hep-ph]].

\end{thebibliography}
\end{document}